\documentclass[twocolumn,showpacs,preprintnumbers,amsmath,amssymb]{revtex4}
\usepackage{graphicx}
\usepackage{dcolumn}
\usepackage{bm}
\usepackage{amsfonts}

\newcommand{\ket}[1]{\mbox{$\mid \! #1 \, \rangle$}}
\newcommand{\bra}[1]{\mbox{$\langle \, #1 \! \mid$}}
\newcommand{\erw}[1]{\langle #1 \rangle}

\newcommand{\dg}[1]{#1 \,^{\circ}}
\newcommand{\bea}{\begin{eqnarray}}
\newcommand{\eea}{\end{eqnarray}}
\newcommand{\beas}{\begin{eqnarray*}}
\newcommand{\eeas}{\end{eqnarray*}}
\newcommand{\WW}{\ensuremath{\mathcal{W}}}

\newcommand{\DD}{\ensuremath{D_{4}^{(2)}}}
\newcommand{\DDv}{\ensuremath{\ket{D_{4}^{(2)}}}}
\newcommand{\DickN}{\ensuremath{\ket{D_{N}^{(M)}}}}
\newcommand{\DickW}{\ensuremath{\ket{D_{N}^{(1)}}}}

\begin{document}
\title{Experimental Observation of Four-Photon Entangled Dicke State with High Fidelity}
\author{N. Kiesel,$^{1,2}$ C. Schmid,$^{1,2}$ G. T\'oth,$^{1,3}$ E. Solano,$^{1,4}$ and H. Weinfurter$^{1,2}$}
\affiliation{ $^{1}$
Max-Planck-Institut f{\"u}r Quantenoptik, Hans-Kopfermann-Strasse 1, D-85748 Garching, Germany \\
$^{2}$Department f\"ur Physik, Ludwig-Maximilians-Universit{\"a}t, D-80797 M{\"u}nchen, Germany \\
$^{3}$ Research Institute for Solid State Physics and Optics, HAS, P.O. Box 49, H-1525 Budapest, Hungary\\
$^{4}$ Secci\'{o}n F\'{\i}sica, Dpto. de Ciencias, Pontificia Universidad Cat\'{o}lica del Per\'{u}, Apartado 1761 Lima, Peru}
\date{\today}
\begin{abstract}
We present the experimental observation of the symmetric four-photon entangled Dicke state with two excitations $|D_{4}^{(2)}\rangle$. A simple experimental set-up allowed quantum state tomography yielding a fidelity as high as $0.844 \pm 0.008$. We study the entanglement persistency of the state using novel witness operators and focus on the demonstration of a remarkable property: depending on the orientation of a measurement on one photon, the remaining three photons are projected into both inequivalent classes of genuine tripartite entanglement, the GHZ and W class. Furthermore, we discuss possible applications of $|D_{4}^{(2)}\rangle$ in quantum communication.
\end{abstract}
\maketitle
Entanglement in bipartite quantum systems is well understood and can be easily quantified. In contrast, multipartite quantum systems offer a much richer structure and various types of entanglement. Thus, crucial questions are how strongly and, in particular, in which way a quantum state is entangled. Therefore, different classifications of multipartite entanglement have been developed \cite{Duer, Verstraete, Chen}.  Further, quantum states with promissing properties and applications have been identified and studied experimentally \cite{W3,Qubyte,Cluster,GHZ,JanWei,Psi4}. The efforts in this direction lead to a deeper understanding of multipartite entanglement and its applications.

In this work, we present a detailed experimental and theoretical
examination of a novel four-photon entangled state: $\DDv$ -- the four-qubit Dicke state with two excitations that is symmetric under all permutations of qubits. Generally, a symmetric $N$-qubit Dicke state \cite{Dicke, Mabu, Enrique} with $M$ excitations is the equally weighted
superposition of all permutations of $N$-qubit product states with $M$ logical $1$'s
and $(N-M)$ logical $0$'s, here denoted by \DickN. Well known examples are the $N$-qubit W states $\ket{W_N}$ (in the present notation \DickW) \cite{Qubyte}. The state $\DDv$, just like $\ket{W_N}$, is highly persistent against photon loss and projective mesurements. In particular we show that, in spite of the impossibility to transform a three photon GHZ type into a W state by local manipulation \cite{Duer}, both can be obtained via a projective measurement of the \emph{same} photon in the state $\DDv$. We study these properties in a simple experimental scheme which allowed the observation of the state with about 60 four-fold coincidences per minute.
For characterization we use quantum state tomography and apply novel witness
operators. Finally we shortly discuss possible applications of the state.
\begin{figure}[t]
\includegraphics[width=8.5cm,clip]{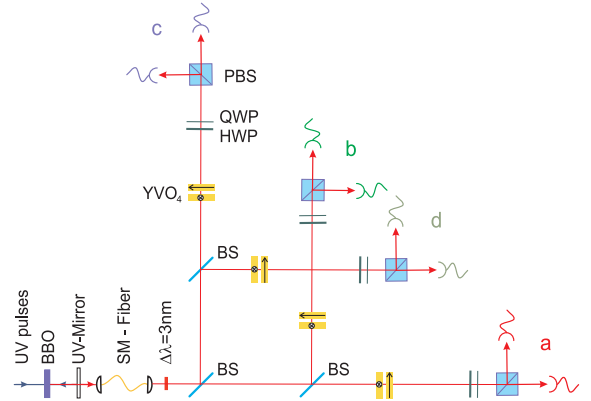}
\caption{Experimental setup for the analysis of the four-photon
polarization-entangled state $\DDv$. It is observed after the
symmetric distribution of four photons onto the spatial
modes a,b,c and d via non-polarizing beam splitters (BS). The photons are
obtained from type-II collinear spontaneous parametric down conversion (SPDC) in a 2~mm $\beta$-Barium
Borate (BBO) crystal pumped by 600~mW UV-pulses. The phases between the
four output modes are set via pairs of birefringent Ytrium-Vanadat-crystals (YVO$_4$). Half and
quarter wave plates (HWP, QWP) together with polarizing beam
splitters (PBS) are used for the polarization analysis.}
\label{fig:setup}
\end{figure}

The state $\DDv$ has the form:
\begin{eqnarray}\notag
\ket{\DD}=\frac{1}{\sqrt{6}}\kern-.5em&(&\kern-.5em\ket{HHVV}+\ket{HVHV}+\ket{VHHV}+\\
\kern-.5em&&\kern-.5em\ket{HVVH}+\ket{VHVH}+\ket{VVHH}) \label{eq:W2}
\end{eqnarray}
with, e.g.,
$\ket{VVHH}=\ket{V}_a\otimes\ket{V}_b\otimes\ket{H}_c\otimes\ket{H}_d$,
where $\ket{H}$ and $\ket{V}$ denote linear horizontal ($H$) and
vertical ($V$) polarization of a photon in the spatial modes
($a,b,c,d$) (Fig. \ref{fig:setup}). Evidently, this is a
superposition of the six possibilities to distribute two horizontally and two vertically
polarized photons into four modes. Accordingly, we create four indistinguishable photons with appropriate polarizations in one spatial mode and distribute them with polarization independent beam splitters (BS) (Fig.~\ref{fig:setup}) \cite{Yamamoto}.
If one photon is detected in each of the four output modes we observe the state $\DDv$.

\begin{figure}[t]
\includegraphics[width=8.5cm,clip]{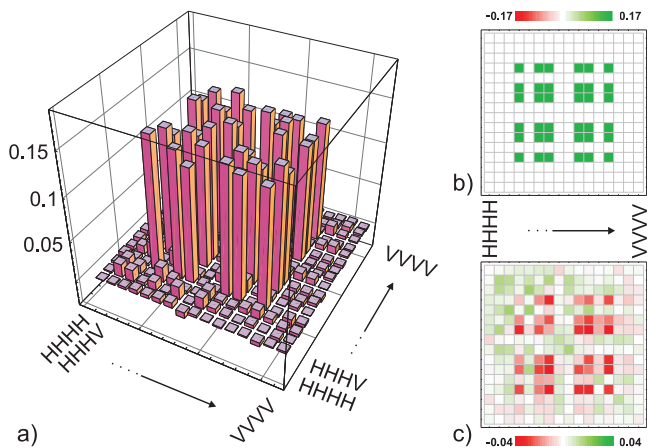}
\caption{(a) Real part of the density matrix $\rho_{\textrm{fit}}$ derived from the observed data, (b) density plot of the ideal state $\rho_{\DD}$ and, for comparison, (c) the difference between the matrices. Noise on the real and imaginary part is comparable.} \label{fig:4qubit}
\end{figure}
As source of the four photons we use the
second order emission of collinear type II spontaneous
parametric down conversion (SPDC). UV pulses with a central wavelength of 390~nm and an
average power of about 600~mW from a frequency-doubled mode-locked
Ti:sapphire laser (pulse length $\approx$130~fs) are used to pump a 2~mm
thick BBO ($\beta$-Barium Borate, type-II) crystal. This results
in two horizontally and two vertically polarized photons with
the same wavelength. Dichroic UV-mirrors serve to separate the
UV-pump beam from the down conversion emission. A half-wave plate
together with a 1~mm thick BBO crystal compensates walk-off effects (not shown in Fig.~\ref{fig:setup}).
Coupling the four photons into a single mode fiber exactly defines
the spatial mode. The spectral selection is achieved with a narrow
bandwidth interference filter ($\Delta\lambda=3$~nm) at the
output of the fiber.
Birefringence in the non-polarizing beam splitter cubes (BS) is compensated with pairs of perpendicularly oriented $200~\mu$m thick
birefringent YVO$_4$ crystals in each of the four modes. Altogether, the setup is stable over several days and mainly limited
by disalignment effects in the pump laser system affecting rather
the count rate than the quality of the state.

Polarization analysis is performed in all of the four outputs. 
For each mode we choose the analysis direction with half (HWP) and quarter waveplates (QWP) and detect the photons with the corresponding orthogonal polarizations in the outputs of polarizing beam splitters using single photon detectors (Si-APD).
The detected signals are fed into a multi-channel coincidence unit
which allows to simultaneously register any possible coincidence between the inputs. The rates for each of the 16
characteristic four-fold coincidences have to be corrected for the
different detection efficiencies in the output modes. The errors
on all quantities are deduced from propagated Poissonian counting
statistics of the raw detection events and the independently
measured detection efficiencies.
\begin{figure*}[t]
\includegraphics[width=\textwidth,clip]{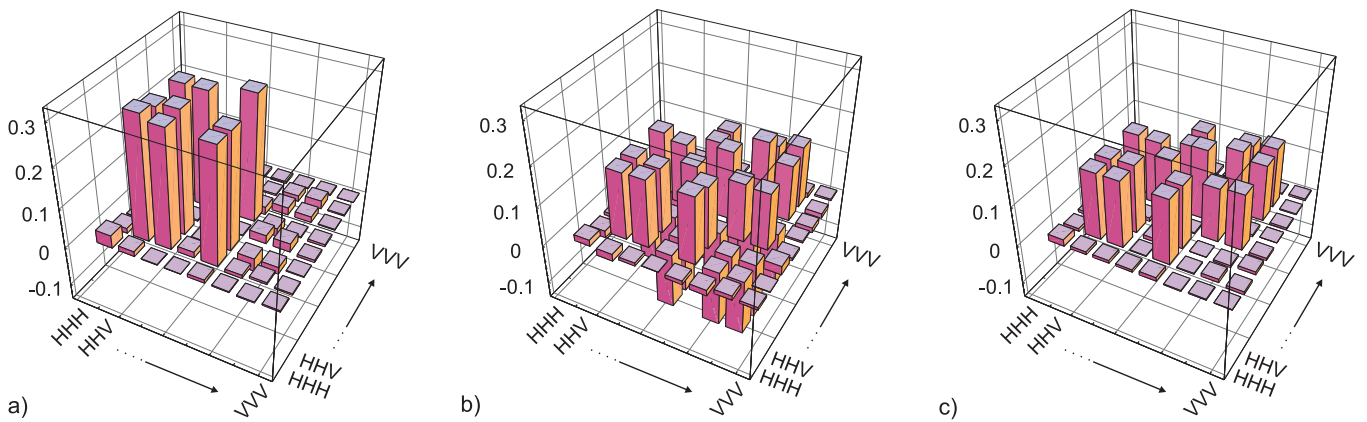}
\caption{Real parts of density matrices for (a) W state after
projection of photon d onto $\ket{V}$, (b) G state after
projection of photon d onto $\ket{-}$, (c) residual state after
loss of photon d. The imaginary parts consist of noise only,
comparable to the noise in the real part.} \label{fig:3qubit}
\end{figure*}

To analyze the observed state we first determine its density matrix. For this purpose we measure the correlations
Tr$\big[\rho_{\textrm{exp}}(\sigma_{i}\otimes\sigma_{j}\otimes\sigma_{k}\otimes\sigma_{l})\big]$
with $i,j,k,l \in \{0,x,y,z\}$, where $\sigma_i$ are the Pauli spin operators. These 256 values can be derived from
the 81 settings of all combinations for analyzing each qubit in one of the three bases: ($H$/$V$), $(\dg{\pm45})$ and ($L$/$R$), where
$\ket{\dg{\pm45}}=1/\sqrt{2}(\ket{H}\pm \ket{V})$ and $\ket{L/R}=1/\sqrt{2}(\ket{H}\pm i\ket{V})$. The measurement was running 35 hours with a count rate of about 60 four-fold coincidences per minute. The obtained data completely characterizes the observed state ($\rho_{\textrm{exp}}$) up to the mentioned, mainly statistical, errors. We use a maximum likelihood approach \cite{MoQ} to estimate a corresponding physical density matrix $\rho_{\textrm{fit}}$. The real part of $\rho_{\textrm{fit}}$ is depicted in Fig.~\ref{fig:4qubit}a. The characteristic structure of the ideal state $\rho_{\DD}$ [Fig.~\ref{fig:4qubit}b] is clearly visible. For comparison Fig.~\ref{fig:4qubit}c shows the differences between the matrices. The noise is about the same in the imaginary part and is mainly caused by higher order emissions and imperfect compensation of the birefringence of fiber and beam splitters. The major difference is in the off diagonal elements: due to the finite spectral bandwidth the coherence gets reduced. This can be improved with narrower filters but at the expense of lower count rates.

To test whether we indeed observe genuine four-partite entanglement we use the generic form of the witness operator $\WW_{\textrm{g}}$ \cite{OldWitness}. The corresponding expectation value depends directly on the observed fidelity: $Tr(\WW_{\textrm{g}} \rho_{\textrm{exp}})=\frac{2}{3}-F_{\textrm{exp}}=-0.177 \pm 0.008$ \cite{explain} and is positive for all biseparable states. In principle, 21 measurement settings, instead of the experimentally expensive complete tomography, are sufficient to determine this value. 

For $\DDv$ one can, however, construct an entanglement witness that is much more efficient. Utilizing the high symmetry of this state, genuine four-partite entanglement can be detected with only two settings via a measurement of the collective spin squared in x- and y- direction ($\erw{J_x^2}$ and $\erw{J_y^2}$). For biseparable states it can be proved that \cite{stabwit2,unpublished}
\begin{equation}
\langle \WW_4^{s} \rangle=\langle J_x^2 \rangle+\langle J_y^2\rangle \le
7/2+\sqrt{3}\approx 5.23, \label{eq:Jxy4}
\end{equation}
where $J_{x/y}=1/2\sum_k \sigma_{x/y}^k$
with e.g.,
 $\sigma_x^3=\openone \otimes \openone \otimes \sigma_x \otimes \openone$.
This can be interpreted also by rewriting $\erw{J_x^2}+\erw{J_y^2}=\erw{J^2} - \erw{J_z^2}$ where $J=(J_x,J_y,J_z)$. As for symmetric states $\erw{J^2} = N/2(N/2+1)$ our criterion requires $\erw{J_z^2} \geq 5/2-\sqrt{3}$, i.e. the collective spin squared of biseparable symmetric states in any direction cannot be arbitrarily small \cite{Collective}.
For the state $\DDv$, however, $\erw{J_z^2}=0$ and thus the expectation value of the witness operator in Eq.~\ref{eq:Jxy4} reaches the maximum of 6. Via measurement of all photons in ($\pm 45$)- and ($L$/$R$)-basis respectively we find experimentally the value $Tr[\WW_4^{s}\rho_{\textrm{exp}}]=5.58 \pm 0.02$ clearly exceeding the required bound. Multipartite entanglement was, thus, detected by studying only a certain property of the state and can, in principle, even be detected without individual adressing of qubits.

Let us start the investigation of the properties that make $\DDv$ special in comparison with the great variety of other four-qubit entangled states studied so far. The various states show great differences in the residual three-qubit state dependent on the measurement basis and/or result: $\ket{GHZ_4}$ \cite{GHZ} can either still render tripartite GHZ like entanglement or become separable, $\ket{W_4}$ as well, but the tripartite entanglement will always be W type. Entanglement in the cluster state $\ket{C_4}$ \cite{Cluster} cannot be easily destroyed and at least bipartite entanglement remains. However, $\DDv$ and also $\ket{\Psi^{(4)}}$ \cite{Robust} yield genuinely tripartite entangled states independent of the measurment result and basis.

Let us compare the projection of the qubit in mode d onto either $\ket{V}$ or $\ket{-}$ for the state $\DDv$:
\begin{eqnarray}
\notag_d\bra{V}\DD\rangle&=&\frac{1}{\sqrt{3}}(\ket{HHV}+\ket{HVH}+\ket{VHH}),\\\notag
_d\bra{-}\DD\rangle&=&\frac{1}{\sqrt{6}}(\ket{HHV}+\ket{HVH}+\ket{VHH}\\
&&-\ket{HVV}-\ket{VHV}-\ket{VVH}). \label{eq:Project}
\end{eqnarray}
The first is the state $\ket{W_3}$ \cite{W3} and the second one is a so-called G state ($\ket{G_3}$ Ref.~\cite{Gdansk}).
Experimentally we observe these states with fidelities $F_{W_3}
=0.882\pm 0.015$ 
and $F_{G_3}
=0.897\pm 0.019$ 
. Comparable values are observed for measurements of photons in other modes. The real part of the
density matrices of the experimental results are depicted in Fig.~\ref{fig:3qubit}. Density matrix (a) shows the measurement result for the state $\ket{W_3}$. In (b) the observed G state is shown containing the entries for $\ket{W_3}$, its spin-flipped counterpart $\ket{\overline{W_3}}$ and, with the negative sign, the coherence terms between the two. Noise in the imaginary part is comparable to the one in the real part.

The criterion (\ref{eq:Jxy4}) adopted to the three-qubit case, can now be used to detect the tripartite entanglement around $\ket{W_3}$ and
$\ket{G_3}$ with the bound $\langle \WW_3^{s} \rangle=\langle J_x^2 \rangle+\langle J_y^2\rangle \le 2+\sqrt{5}/2\approx 3.12$.
Our measurement results for
$\ket{W_3}$ and $\ket{G_3}$ 
are $Tr\big[\WW_3^{s}
\rho_{G_3}\big]=3.34 \pm 0.03$ 
and $Tr \big[\WW_3^{s}
\rho_{W_3}\big]=3.33 \pm 0.03$ 
 respectively, which proves
both states contain genuine tripartite entanglement.

What kind of tripartite entanglement do we observe? Fascinatingly, this depends on the measurement basis.
While the W state represents the W class, the state $\ket{G_3}$ belongs to the GHZ class.
This is extraordinary: GHZ and W class states can not be transformed
into one another via SLOCC \cite{Duer} and not even by entanglement catalysis
\cite{Mohamed}. $\DDv$, however, can be projected into both classes by a local operation, i.e., 
via a simple von Neumann measurement of one qubit. This also implies that there is no obvious way how to prepare $\DDv$ out of either of those three-qubit states via a 2-qubit interaction with an additional photon, as this would directly give a recipe to transform one class of three-qubit entanglement into the other. 
As the experimentally prepared states are not perfect we also have to test whether the observed state $\ket{G_3}$ is GHZ class.
To do so we construct an entanglement witness from the generic one for pure GHZ states, $\WW_{GHZ_3}=\frac{3}{4}\openone-\ket{GHZ_3}\bra{GHZ_3}$, by applying local filtering operations $\widehat{F}=A \otimes B \otimes C$. The resulting witness operator is then $\WW'=\widehat{F}^\dagger \WW_{GHZ_3}\widehat{F}$ \cite{Otfried, Qubyte}. Here $A,B$ and $C$ are $2\times 2 $ complex matrices determined through numerical optimization to find an optimal witness for the detected state. Note, that $\WW'$ still detects GHZ type entanglement as $\widehat{F}$ is an SLOCC operation. In the measurement GHZ type entanglement is indeed detected with an expectation value of $Tr(\rho_G \WW')=-0.029 \pm 0.023$ proving that the observed state is \emph{not} W class.

Entanglement in $\DDv$ is not only persistent against projective measurements but also against loss of photons. The state $\rho_{\textrm{abc}}$ after tracing out qubit d is an equally weighted mixture of $\ket{W_3}$ and $\ket{\overline{W_3}}$, which is also tripartite entangled Fig.~\ref{fig:3qubit}(c).
Applying witness $\WW_3^s$ we obtain $Tr\big[\WW_3^{s} \rho_{\textrm{abc}}\big]=3.30 \pm 0.01$
, proving clearly the genuine tripartite entanglement. The fidelity
with respect to the expected state is $F_{\textrm{abc}}=0.924 \pm 0.006$, similar values are reached for the loss of the photons in modes a, b and c. We observe the contributions of $\ket{W_3}$ and
$\ket{\overline{W_3}}$, but contrary to the state $\ket{G_3}$ [Fig.~\ref{fig:3qubit}(b)]
there is no coherence between the two.

As we have seen, the loss of one photon results in a three-qubit entangled W class state. Thus, the persistency against the loss of a second photon should also be high \cite{Robust}. It is known that the state $\ket{W_4}$ is the symmetric state with the highest persistency against loss of two photons with respect to entanglement measures like the concurrence \cite{Duer, Mabu}. In contrast, it turns out that for $\DDv$ the remaining two photons have the highest possible maximal singlet fraction \cite{MSF} (MSF$_{\DD}=2/3$, experimentally MSF$_{\textrm{exp}}=0.624 \pm 0.005$). This means that the residual state is as close to a Bell state as possible. It was already pointed out in Refs.~\cite{Robust, MSF} that this is a hint for the applicability of a state in telecloning \cite{telecloning}. Four parties that share the state $\DDv$ can use the quantum correlations in each pair of qubits as a quantum channel for a teleportation protocol. Thus, each party can distribute an input qubit to the other parties with a certain fidelity, which depends on the MSF. Using $\DDv$ as quantum resource this so-called $1\rightarrow3$~telecloning works with the optimal fidelity allowed by the no-cloning theorem. Averaged over arbitrary input states the fidelity is F$^{\textrm{clone}}_{1\rightarrow3}=0.788$ and the optimal so-called covariant cloning fidelity is F$^{\textrm{cov}}_{1\rightarrow3}=0.833$ for all input states on the equatorial plane of the Bloch sphere.

What if the receiving parties decide that one of them should get a perfect version of the input state? Probabilistically this is still possible, if the other two parties abandon their part of the information by a measurement of their qubit in the same direction, say (H/V). In case they find orthogonal measurement outcomes the sender and the only remaining receiver share a Bell state $_{cd}\bra{HV}\DD\rangle=\frac{1}{\sqrt{3}}(\ket{HV}+\ket{VH})=\sqrt{\frac{2}{3}}\ket{\psi^+}_{ab}$. This enables perfect teleportation in 2/3 of the cases and therefore, as each party could be the receiver, an open destination teleportation (ODT) \cite{JanWei}. The experimentally obtained fidelity in this case was $F_{H_cV_d}^{\psi^+}=0.883 \pm 0.028$
. For other measurement directions different Bell states can be obtained. For example for projections onto the ($\pm45$)- and (L/R)-basis we found $F_{+_c-_d}^{\phi^+}=0.721 \ \pm 0.043$ 
and $F_{R_cL_d}^{\phi^-}=0.712 \pm
0.042$
. Note that, in contrast to the deterministic GHZ based ODT protocol, $\DDv$ allows to choose between telecloning and ODT.

Finally, as another possible application, we also note
that $\DDv$ is one of the two symmetric Dicke states which can
be used in certain quantum versions of classical games
\cite{game}. In these models it might offer new game
strategies compared to the commonly used GHZ state.

In conclusion we have presented the experimental observation and
analysis of a quantum state $\DDv$, obtained
with a fidelity of $0.844 \pm 0.008$ and a count rate as high as 60
counts/minute. 
The setup and methods used are generic for observation of symmetric Dicke states with higher photon numbers.
An analysis of the state after projection of one qubit in different bases showed that the two inequivalent classes of genuine tripartite entanglement can be obtained. An optimized entanglement witness served to verify this experimentally. We also show that the possibility to project two photons into a Bell state makes $\DDv$ a resource for an ODT protocol. Further, the state has a high entanglement persistency against loss of two photons. In this case, the singlet fraction of the remaining photons is maximal and from this we infered applicability of the state for quantum telecloning.
We are confident that due to the extraordinary properties of the state more
applications are likely.

We acknowledge the support of this work by the Bavarian High-tech Inintiative, the Deutsche Forschungsgemeinschaft, the European Commission through the EU Projects QUAP and RESQ and the EU Grants MEIF-CT-2003-500183 and MERG-CT-2005-029146, and the National Research Fund of Hungary OTKA (Contract No. T049234).


\begin{thebibliography}{99}
\bibitem{explain} All experimental values presented in the following are calculated from the data directly and not from the fit.
\end{thebibliography}
\end{document}